\documentclass[a4paper,12pt]{article}
\usepackage{type1cm,amsmath,amssymb,color,graphics,amscd,amsfonts,mathrsfs,indentfirst,cite}
\usepackage{epsfig}

\usepackage[dvipdfm,hypertex]{hyperref}

\makeatletter
\@addtoreset{equation}{section}
\renewcommand{\theequation}{\thesection.\@arabic\c@equation}
\makeatother

\makeatletter
\renewcommand\appendix{\par
  \setcounter{section}{0}%
  \setcounter{subsection}{0}%
  \gdef\thesection{Appendix \@Alph\c@section :\!\!\!}
  \renewcommand{\theequation}
  {\Alph{section}.\arabic{equation}}
}
\makeatother

\newcommand{\dd}{{\rm d}}

\makeatletter
\renewcommand\section{\@startsection {section}{1}{\z@}%
                                   {-3.5ex \@plus -1ex \@minus -.2ex}
                                   {2.3ex \@plus.2ex}%
                                   {\normalfont\large\bfseries}}
\renewcommand\subsection{\@startsection{subsection}{2}{\z@}%
                                     {-3.25ex\@plus -1ex \@minus -.2ex}%
                                     {1.5ex \@plus .2ex}%
                                    {\normalfont\bfseries}}
\makeatother

\begin{document}

\begin{titlepage}
\thispagestyle{empty}

\vspace*{-15mm}   
\baselineskip 10pt   
\begin{flushright}   
\begin{tabular}{r}    
{\tt HRI/ST/1010}\\
{\tt PUPT-2354}\\
October 2010
\end{tabular}   
\end{flushright}   
\baselineskip 24pt   
\vglue 10mm   

\begin{center}
{\Large\bf
 New Near Horizon Limit in Kerr/CFT
}

\vspace{10mm}   

\baselineskip 18pt   

\renewcommand{\thefootnote}{\fnsymbol{footnote}}

Yoshinori~Matsuo$^a$\footnote[2]{ymatsuo@hri.res.in} and
Tatsuma~Nishioka$^b$\footnote[3]{nishioka@princeton.edu}

\renewcommand{\thefootnote}{\arabic{footnote}}
 
\vspace{10mm}   

{\it  
${}^a$ Harish-Chandra Research Institute, 
 Chhatnag Road, Jhusi, Allahabad 211 019, India \\

\bigskip

${}^b$ Department of Physics, Princeton University, Princeton, NJ 08544, USA\\
}
  
\vspace{25mm}   

\bigskip

\end{center}

\begin{abstract}
The extremal Kerr black hole with the angular momentum $J$ 
is conjectured to be dual to CFT with central charges $c_L = c_R = 12J$. 
However, the central charge in the right sector remains to be explicitly derived so far.
In order to investigate this issue, we introduce new near horizon limits of 
(near) extremal Kerr and five-dimensional Myers-Perry black holes.
We obtain Virasoro algebras as asymptotic symmetries and
calculate the central charges associated with them.
One of them is equivalent to that of the previous studies, and
the other is non-zero, but still the order of near extremal parameter.
Redefining the algebras to take the standard form, we obtain a finite value
as expected by the Kerr/CFT correspondence.
\end{abstract}

\end{titlepage}

\baselineskip 18pt   
\section{Introduction and overview}\label{sec:Intro}

The origin of the black hole entropy has been of great interest in the past
few decades and 
remains to be fully understood. 
In string theory, some of black holes and branes can be 
described in terms of the microscopic degrees of freedom, 
and the entropy is reproduced by counting their microstates. 
In the case of anti-de Sitter (AdS) spacetime, 
these microstates are described by the conformal field theory (CFT). 
In this case, a lot of information can be 
obtained by using symmetries. 
In the case of AdS$_3$, the Bekenstein-Hawking entropy 
of the BTZ black hole
was accounted without specifying details of CFT. 
The asymptotic symmetry of the geometry is identified 
with the conformal symmetry of the dual field theory and the entropy is 
calculated by using the Cardy formula \cite{bh,Strominger:1997eq}. 

Recently, it was conjectured that 
the extremal Kerr black hole in four dimensions corresponds to a two-dimensional CFT \cite{ghss}. 
They investigated its near horizon geometry 
which has the $SL(2,\mathbb R)\times U(1)$ isometry \cite{baho}. 
The $U(1)$ symmetry is enhanced to an asymptotic symmetry and 
identified with the Virasoro algebra for the chiral half of CFT. 
The Bekenstein-Hawking entropy is reproduced 
by using the Cardy formula in the extremal case. 
This is called as the Kerr/CFT correspondence 
and this Virasoro algebra is regarded as that of left mover. 
This analysis is generalized to many cases \cite{many}\footnote{See also \cite{Horowitz:2007xq,Dias:2007nj} for related works.} and
it is known that another Virasoro algebra can be obtained 
from extending the $SL(2,\mathbb R)$ part of the isometry \cite{mty, cl}.
This symmetry is considered as that of right mover
which describes the non-extremal excitation  
from the extremality. 
Therefore, the extremal Kerr black hole with an angular momentum $J$
is expected to be dual to a non-chiral CFT with $c_L=c_R=12J$.
The recent discussion about the hidden conformal symmetry of the non-extremal Kerr black hole
also provides another evidence for this correspondence \cite{cms}.

However, the asymptotic Virasoro algebras can be found 
only in the near horizon limit of (near) extremal black holes. 
Moreover, the central charge for right mover is zero when it is 
evaluated by the conventional method.
Therefore, a cut-off was introduced into the spacetime 
to take the near-extremal correction into account, and
it turned out to be not finite but infinitesimally small \cite{mty}.
Although an indirect argument is presented in \cite{cl},
it is fair to say that the central charge $c_R=12J$ 
has not been explicitly derived so far.

In this paper, we investigate this issue in more detail.
First, we study the five-dimensional extremal Myers-Perry black hole
whose near horizon geometry includes
AdS$_3$ if it has only a single spin \cite{baho}\footnote{The 
structure of AdS$_3$ sometimes appears 
in certain limits of extremal rotating black holes. 
It has been recently studied to investigate
 the Kerr/CFT correspondence in \cite{Guica:2010ej,Azeyanagi:2010pw}.}. 
In this case, we can find 
dual two-dimensional CFT with central charges $c_L=c_R$
similar to the one given by Brown and Henneaux \cite{bh}. 
The CFT lives on light-cone coordinates 
which consist of the time and angular directions. 
On the other hand, we know the Kerr/CFT description where dual CFT is defined on
the time and angular directions with central charges $c_L\neq c_R$. 
The coordinates in Kerr/CFT description are 
almost equivalent to the light-cone coordinates 
near the horizon up to a scaling factor.
This scaling factor makes central charges different from each other. 

To resolve this discrepancy, 
we reconsider the definitions of the algebras, 
central charges and temperatures. 
Usually, the Frolov-Thorne temperature is 
changed under the rescaling of the coordinates, 
while the central charges are fixed 
because it represents a number of degrees of freedom. 
If we use these definitions of the central charges and temperatures, 
the Cardy formula gives scale dependent quantity. 
However, the entropy should be independent of the scale 
since it counts a number of states. 
Therefore we need to introduce ``covariant'' 
definition of the central charge to make the Cardy formula invariant under 
the scaling. 
%
The covariant central charge $c^{(cov)}$ is related to 
the scale-invariant central charge $c$ as $c^{(cov)} = \beta c$, 
where $\beta$ is the period of the coordinate. 
Then, the covariant central charge depends on the scaling factor 
via the period $\beta$. 
In the Kerr/CFT correspondence, 
the Virasoro algebra for right mover has been 
studied by using the quasi-local charges \cite{by,bk}. 
In these studies generators are defined in the scale covariant form, 
and the central charge depends on the definition of 
the time coordinates in the near horizon geometry. 
In the near horizon limit, these covariant central charge 
for left and right movers are not generally equal to each other, 
$c_L^{(cov)}\neq c_R^{(cov)}$. 
In this paper, we show that the central charges 
satisfy the relation $c_L = c_R$ even in Kerr/CFT, 
if we use the scale-invariant definition of $c$. 

We also consider general five-dimensional Myers-Perry black hole 
with two rotations,
where the near horizon geometry is AdS$_2$, not AdS$_3$.
Thus there is no natural light-cone coordinates, but we introduce 
new coordinates which agree with the light-cone coordinates 
in the single rotation limit. 
The resulting near horizon geometry is the same 
as the usual one once replacing the time and angular directions 
with the light-cone directions.
We evaluate the central charge for our new near horizon geometry 
on the timeslice with respect to the original time. 
%
By using this coordinates, we can evaluate
the central charge associated with 
the Virasoro algebra for the right mover 
using the conventional covariant phase space method \cite{bb,bc} 
and obtain the scale-invariant central charges satisfying $c_L = c_R$. 
We apply this analysis to the (four-dimensional) Kerr black hole 
and obtain the expected value of $c_L = c_R = 12J$. 

It is worth noting that the combinations 
of coordinates in our new near horizon limits 
are the same as those in the analyses of 
the hidden conformal symmetry \cite{cms}. 
In these studies our covariant central charges 
are the equivalent to the invariant one. 
Then, the Cardy formula can be used with 
the Frolov-Thorne temperature and 
these central charges to calculate the entropy. 
Therefore the central charges $c_L = c_R = 12J$ is naturally obtained 
in the original coordinates before taking the near horizon limit.

%

This paper is organized as follows. 
In Section~\ref{sec:Review}, 
we introduce the five-dimensional Myers-Perry black hole 
and its near horizon limit. Then, we review on the 
Kerr/CFT correspondence for left mover. 
In Section~\ref{sec:QL}, 
we show the calculation of the central charge for right mover
following the previous study \cite{mty}. 
We introduce a cut-off to evaluate the central charge. 
In Section~\ref{sec:AdS}, we consider 
the special case in which the near horizon geometry 
has the structure of AdS$_3$. 
Then, we discuss the relation between this special case and 
the general case described in Section~\ref{sec:Review} and \ref{sec:QL}. 
In Section~\ref{sec:NewNH}, we introduce 
a new definition of the near horizon limit. 
By using this definition, we calculate the 
central charge for right mover without 
introducing the cut-off into the spacetime. 
In Section~\ref{sec:Kerr}, 
we apply the new definition of 
the near horizon limit to the Kerr black hole, 
and calculate the central charge. 
Section~\ref{sec:Concl} is devoted to the conclusion.

\section{Kerr/CFT correspondence in 5D Myers-Perry black hole}\label{sec:Review}

In this section, we describe 
the five-dimensional Myers-Perry black hole and 
briefly review on the Kerr/CFT correspondence for it. 
The metric is expressed by using 
the Boyer-Lindquist coordinates as 
\begin{align}
 ds^2 
 &= 
 - \frac{\Delta}{r^2\rho^2}
 \left(dt -a \sin^2\theta\,d\phi - b \cos^2\theta\,d\psi\right)^2 
 \notag\\
 &\quad 
 + \frac{\sin^2\theta}{\rho^2}\left[(r^2+a^2)d\phi -a\,dt \right]^2
 + \frac{\cos^2\theta}{\rho^2}\left[(r^2+b^2)d\psi -b\,dt \right]^2
 \notag\\
 &\quad
 +\frac{1}{r^2\rho^2}
 \left[
 b(r^2+a^2)\sin^2\theta\,d\phi +a(r^2+b^2)\cos^2\theta\,d\psi -ab\,dt
 \right]^2 
 \notag\\
 &\quad
 + \frac{r^2\rho^2}{\Delta^2} dr^2 + \rho^2 d\theta^2 \ , 
\end{align}
where $\Delta$ and $\rho^2$ are given by 
\begin{align}
 \Delta 
 &= 
 (r^2+a^2)(r^2+b^2) -\mu r^2 \ , 
 & 
 \rho^2 
 &= 
 r^2 + a^2\cos^2\theta + b^2\sin^2\theta \ . 
\end{align}
In five-dimensional spacetime, we have two independent 
angular momenta in $\phi$- and $\psi$-directions, respectively. 
Hence, this geometry has three parameters $\mu$, $a$, $b$, 
which are related to the ADM mass $M_{\text{ADM}}$ 
and two angular momenta $J_\phi$ and $J_\psi$:
\begin{align}
 M_\text{ADM} &= \frac{3\pi\mu}{8G_N} \ , & 
 J_\phi &= \frac{\pi\mu a}{4G_N} \ , & 
 J_\psi &= \frac{\pi\mu b}{4G_N} \ ,  
\end{align}
where $G_N$ is the Newton constant. 
The outer and inner horizons are located at 
the radii $r_+$ and $r_-$, which are expressed as 
\begin{equation}
 r_\pm^2 
  = 
  \frac{1}{2}
  \left(
   \mu-a^2-b^2 \pm\sqrt{\left(\mu-(a+b)^2\right)(\mu-(a-b)^2)}
  \right) \ , 
\end{equation}
and the angular velocities on the (outer) horizon are 
\begin{align}
 \Omega_\phi &= \frac{a}{r_+^2+a^2} \ ,& 
 \Omega_\psi &= \frac{b}{r_+^2+b^2} \ .  
\end{align}
The Hawking temperature $T_H$ and 
the Bekenstein-Hawking entropy $S$ 
are given by 
\begin{align}
 T_H &= \frac{r_+^2 - r_-^2}{2\pi\mu r_+} \ , & 
 S &= \frac{\pi^2}{2G_N}\mu r_+ \ , 
\end{align}
respectively. 

Now, we consider the near horizon limit of this geometry \cite{baho}. 
We focus on the near-extremal case 
in which the non-extremality is infinitesimally small. 
We define this non-extremality parameter by 
\begin{align}
 \mu &= \mu_0 (1+\epsilon^2\hat\mu) \ , & 
 \mu_0 = (a+b)^2 \ ,  \label{NEcond}
\end{align}
where the extremal condition is given by $\mu = \mu_0$ 
and $\hat\mu$ parametrizes the non-extremality. 
We introduce the near horizon coordinates as 
\begin{subequations}\label{NHCoord}
 \begin{align}
  t &= \epsilon^{-1} \frac{\sqrt{\mu_0}}{2} \hat t \ , & 
  r &= r_0 + \epsilon\frac{\sqrt{\mu_0}}{2}\hat r \ , 
  \\
  \phi &= \hat \phi + \frac{a}{r_0^2+a^2}t \ , &
  \psi &= \hat \psi + \frac{b}{r_0^2+b^2}t \ ,
 \end{align}
\end{subequations}
where $r_0$ is the horizon radius 
in the extremal case and given by 
\begin{align}
 r_0^2 &= ab \ .  
\end{align}
Then, the near horizon limit is obtained 
as the $\epsilon\to 0$ limit. 
In this limit, the metric becomes 
\begin{align}
 ds^2 
 &= 
 - \frac{\rho_0^2\hat\Delta}{4} d\hat t^2 
 + \frac{\rho_0^2}{4\hat\Delta}d\hat r^2 
 + \rho_0^2 d\theta^2 
 \notag\\
 &\quad
 +\frac{a^2\mu_0\sin^2\theta}{\rho_0^2}
 \left(
 d\hat\phi + k_{\hat\phi}\hat r d\hat t
 \right)^2
 +\frac{b^2\mu_0\cos^2\theta}{\rho_0^2}
 \left(
 d\hat\psi + k_{\hat\psi}\hat r d\hat t
 \right)^2
 \notag\\
 &\quad 
 +\frac{\mu_0r_0^2}{\rho_0^2}
 \left[
 \sin^2\theta\left(
 d\hat\phi + k_{\hat\phi}\hat r d\hat t
 \right)
 +\cos^2\theta\left(
 d\hat\psi + k_{\hat\psi}\hat r d\hat t
 \right)
 \right]^2  \ ,\label{NHmetric}
\end{align}
where 
\begin{align}
 \hat \Delta &= (\hat r^2 - \hat \mu) \ , & 
 \rho_0 &= r_0^2 + a^2\cos^2\theta + b^2\sin^2\theta \ ,
 \\
 k_{\hat\phi} &= \frac{1}{2}\sqrt{\frac{b}{a}} \ , &
 k_{\hat\psi} &= \frac{1}{2}\sqrt{\frac{a}{b}} \ . 
\end{align}

A correspondence between this near horizon geometry 
and its dual CFT is studied in \cite{lmp}. 
It was shown that the asymptotic symmetry 
gives the Virasoro algebra for left mover 
if we take an appropriate boundary condition. 
The asymptotic symmetry is defined as a symmetry 
which preserves a boundary condition: 
\begin{equation}
 \pounds_\xi \left(g_{\mu\nu}+\mathcal O(\chi_{\mu\nu})\right)
  = \mathcal O(\chi_{\mu\nu}) \ , 
\end{equation}
where $\chi_{\mu\nu}$ defines the  boundary condition. 
If the geometry takes the following form: 
\begin{align}
 ds^2 
 &= 
 f_0(\theta) \left(-\hat r^2 d\hat t^2 +\frac{d\hat r^2}{\hat r^2}\right)
 + \gamma_{ij}(\theta)
 \left(dx^i+k^i r dt\right)\left(dx^j+k^j r dt\right)
 + f_\theta(\theta) d\theta^2 \ , \label{GeneralNH}
\end{align}
where $x^1 = \hat\phi$ and $x^2 = \hat\psi$, 
we can obtain the following asymptotic symmetry groups
\begin{subequations}\label{AsymptKillingLBoth}
 \begin{align}
 \xi^{(\hat\phi)} 
 &= 
 \epsilon_{\hat\phi}(\hat\phi)\partial_{\hat\phi} 
 - \hat r \epsilon_{\hat\phi}'(\hat\phi) \partial_{\hat r} \ , 
 \\
 \xi^{(\hat\psi)} 
 &= 
 \epsilon_{\hat\psi}(\hat\psi)\partial_{\hat\psi} 
 - \hat r \epsilon_{\hat\psi}'(\hat\psi) \partial_{\hat r} \ , 
 \end{align}
\end{subequations}
by taking appropriate boundary conditions, respectively. 
These vectors form the Virasoro algebra as we will see below. 
We define $\xi_n^{(\hat\phi)}$ and $\xi_n^{(\hat\psi)}$ 
by $\xi^{(\hat\phi)}$ and $\xi^{(\hat\psi)}$ 
with 
\begin{align}
 \epsilon_{\hat\phi}(\hat\phi) &= e^{in\hat\phi} \ , & 
 \epsilon_{\hat\psi}(\hat\psi) &= e^{in\hat\psi} \ , 
\end{align}
respectively. 
Then these  vectors obey 
\begin{align}
 [\xi_n^{(\hat\phi)}, \xi_m^{(\hat\phi)}] 
 &= -i(n-m)\xi_{n+m}^{(\hat\phi)} \ , & 
 [\xi_n^{(\hat\psi)}, \xi_m^{(\hat\psi)}] 
 &= -i(n-m)\xi_{n+m}^{(\hat\psi)} \ , 
\end{align}
respectively. 

In order to calculate the central charge, 
we consider the conserved charge associated with the Virasoro algebras. 
A definition of the conserved charge is 
given by \cite{bb,bc}. 
The conserved charge is expressed in terms of the 
background metric $\bar g_{\mu\nu}$ and 
its small perturbation $h_{\mu\nu}$, and given by 
\begin{equation}\label{conservedcharge}
 Q_\xi[h] 
  = \frac{1}{8\pi G_N}
  \!\int_{\partial\Sigma}
  k_{\xi}[h,\bar g] \ , 
\end{equation}
where $\Sigma$ is a timeslice and 
the integration is taken over its boundary $\partial\Sigma$. 
The three-form $k_\xi$ is defined by 
\begin{equation}
 k_{\xi}[h,\bar g] 
 = 
 \tilde k_{\xi}^{\mu\nu}[h,\bar g]\, 
 \left(\dd^3 x\right)_{\mu\nu} \ , 
\end{equation}
where $\dd^3x$ is the Hodge dual of 
the two-form $\dd x^\mu\wedge\dd x^\nu$, 
and the two-form $\tilde k_\xi[h,\bar g]$ is given by 
\begin{align}
 \tilde k_{\xi}^{\mu\nu}[h,\bar g]
 = \frac{1}{2}\Bigl[
& 
 \xi^\mu D^\nu h 
 - \xi^\mu D_\lambda h^{\lambda\nu} 
 + \left(D^\mu h^{\nu\lambda}\right)\xi_\lambda 
 + \frac{1}{2} h D^\mu \xi^\nu 
 \notag\\
&
 - h^{\mu\lambda}D_\lambda\xi^\nu 
 + \frac{1}{2}h^{\mu\lambda}
 \left(D^\nu\xi_\lambda+D_\lambda\xi^\nu\right) 
 - (\mu\leftrightarrow\nu) \Bigr] \ . 
\end{align}
The central charge $c$ can be read off from the 
anomalous transformation of the charge: 
\begin{equation}
 \frac{1}{8\pi G_N}
  \!\int_{\partial\Sigma}
  k_{\xi_n}[\pounds_{\xi_m}\bar g,\bar g] 
  = \delta_{n+m,0}\,n^3\frac{c}{12} \ . 
\end{equation}
For the metric \eqref{GeneralNH}, 
and asymptotic symmetry groups \eqref{AsymptKillingLBoth}, 
we obtain 
\begin{align}
 c_{\hat\phi} &=  \frac{6 \pi k_{\hat\phi}}{G_N} 
  \int d\theta\sqrt{\gamma(\theta) f_\theta(\theta)} \ , & 
 c_{\hat\psi} &=  \frac{6 \pi k_{\hat\psi}}{G_N} 
  \int d\theta\sqrt{\gamma(\theta) f_\theta(\theta)} \ .  
\end{align}
By using the explicit form of the metric \eqref{NHmetric}, 
we obtain 
\begin{align}
 c_{\hat\phi} &= \frac{3\pi b\mu_0}{2G_N} \ , & 
 c_{\hat\psi} &= \frac{3\pi a\mu_0}{2G_N} \ . \label{CentralChargePsi}
\end{align}
Since the Frolov-Thorne temperatures are given by 
\begin{align}
 T_{\hat\phi} &= \frac{(a+b)(r_+ + r_-)}{2\pi(r_+^2+b^2)} 
 \to \frac{r_0}{\pi b} \  , 
 \\
 T_{\hat\psi} &= \frac{(a+b)(r_+ + r_-)}{2\pi(r_+^2+a^2)} 
 \to \frac{r_0}{\pi a} \ , 
\end{align}
the Cardy formula reproduces 
the Bekenstein-Hawking entropy at the extremality: 
\begin{equation}
 S 
  = \frac{\pi^2}{3}c_{\hat\phi} T_{\hat\phi} 
  = \frac{\pi^2}{3}c_{\hat\psi} T_{\hat\psi} 
  = \frac{\pi^2}{2}\mu_0 r_0 \ . 
\end{equation}
It should be noted that both CFTs corresponding to each asymptotic symmetry can reproduce the entropy. 

\section{Near-extremal correction for 5D Myers-Perry black hole}\label{sec:QL}

In this section, we consider 
the Kerr/CFT correspondence for the right mover 
in the five-dimensional Myers-Perry black hole. 
For the Kerr black hole, 
the asymptotic symmetry for the right mover 
is obtained by introducing a different boundary condition to 
that for the left mover \cite{mty}. 
For the five-dimensional Myers-Perry black hole, 
we can obtain a similar asymptotic symmetry group
with an analogous boundary condition. 
For metrics which has the form of \eqref{GeneralNH}, 
we impose the following boundary condition: 
\begin{equation}
 \mathcal O(\chi_{\mu\nu}) = 
  \bordermatrix{
  & \hat t & \hat r & \hat\phi & \hat\psi & \theta \cr
  \hat t 
  & \mathcal O(r^{0}) 
  & \mathcal O(r^{-3})
  & \mathcal O(r^{-2}) 
  & \mathcal O(r^{-2}) 
  & \mathcal O(r^{-3}) 
  \cr 
  \hat r 
  & 
  & \mathcal O(r^{-4})
  & \mathcal O(r^{-3})
  & \mathcal O(r^{-3})
  & \mathcal O(r^{-4}) 
  \cr 
  \hat \phi 
  & 
  & 
  & \mathcal O(r^{-2})
  & \mathcal O(r^{-2})
  & \mathcal O(r^{-3}) 
  \cr 
  \hat \psi 
  & 
  & 
  &
  & \mathcal O(r^{-2})
  & \mathcal O(r^{-3})
  \cr
  \theta
  &
  & 
  & 
  & 
  & \mathcal O(r^{-3})
  }\ . \label{AsymptCondR}
\end{equation}
Then, the asymptotic symmetry group 
\begin{align}
 \xi &= 
  \Big(
   \epsilon_\xi(\hat t) + \frac{\epsilon_\xi''(\hat t)}{2\hat r^2}
  \Big)
  \partial_{\hat t} 
  + 
  \Big(
   - \hat r \epsilon_\xi'(\hat t) 
 + \frac{\epsilon_\xi'''(\hat t)}{2\hat r}
  \Big)
  \partial_{\hat r} 
 \notag\\
 &\quad
  + 
  \Big(
   C_{\hat\phi} - \frac{k_{\hat\phi} \epsilon_\xi''(\hat t)}{\hat r}
  \Big)
  \partial_{\hat\phi}
  + 
  \Big(
   C_{\hat\psi} - \frac{k_{\hat\psi} \epsilon_\xi''(\hat t)}{\hat r}
  \Big)
  \partial_{\hat\psi} 
  + 
  \mathcal O(\hat r^{-3})\ , 
 \label{AsymptKillingR}
\end{align}
preserves the boundary condition for the metric 
\begin{equation}
 g_{\mu\nu} + \mathcal O(\chi_{\mu\nu})
 \to g_{\mu\nu} + \mathcal O(\chi_{\mu\nu})\ . 
\end{equation}

For the asymptotic symmetry group \eqref{AsymptKillingR}
the central charge vanishes, and hence, 
we have to introduce a cut-off. 
Here, we calculate the central charge 
by using the quasi-local charge \cite{by,bk}.
This charge is defined as 
an integration of the surface energy-momentum tensor 
on the boundary. 
The surface energy-momentum tensor is 
given by the conjugate momentum of the induced metric 
on the surface, and can be expressed in terms of 
the extrinsic curvature as 
\begin{equation}
 T^{\mu\nu} 
= \frac{2}{\sqrt{-\gamma}}\,\pi^{\mu\nu} 
= \frac{1}{8\pi G_N} 
  \left(K^{\mu\nu}-\gamma^{\mu\nu}K\right) \ , 
\end{equation}
where $\gamma_{\mu\nu}$ is the induced metric on the boundary 
and $\pi_{\mu\nu}$ is its conjugate momentum, and 
$K_{\mu\nu}$ is the extrinsic curvature. 
Here, we consider the metric with small perturbation $h_{\mu\nu}$, 
and take the difference of the surface energy-momentum tensor 
from that of the background $\bar g_{\mu\nu}$:%
\footnote{
Instead of taking difference from 
the background, we can introduce a 
counter term such that the charge $Q_\xi$ becomes finite (see Appendix~A). 
} 
\begin{equation}
 \tau^{\mu\nu}[h] = 
  T^{\mu\nu}\Bigr|_{g=\bar g + h} - T^{\mu\nu}\Bigr|_{g=\bar g} \ . 
  \label{SurfaceEMTensor}
\end{equation}
Then, the quasi-local charge is defined as 
\begin{equation}
 Q^{\rm QL}_\xi = \!\int_{\partial\Sigma}\!\! \dd^2 x\sqrt{\sigma}\,
  u^\mu \tau_{\mu\nu} \xi^\nu\ . \label{QLCharge}
\end{equation}
where $u^\mu$ is a timelike unit normal to a timeslice $\Sigma$
and $\sigma$ is an induced metric on the timeslice at the boundary $\partial\Sigma$. 
This quasi-local charge corresponds to 
the energy-momentum tensor for the right mover in CFT as, 
\begin{equation}
 Q_\xi^\mathrm{QL} \sim \bar T(\bar z)\bar\epsilon(\bar z) \ . 
\end{equation}
The central extension can be read off from 
the anomalous transformation of this charge: 
\begin{equation}
 \delta_\xi Q_\zeta^\mathrm{QL} = \int_{\partial\Sigma}\!\! \dd^2 x\sqrt{\sigma}\,
  u^\mu \tau_{\mu\nu}[\pounds_\xi \bar g] \zeta^\nu\ . 
\end{equation}
For two-dimensional CFT, 
the central extension of the Virasoro algebra 
can be read off from the anomalous transformation 
of the energy-momentum tensor. 
In an analogous fashion to this, 
we estimate the central extension from the 
anomalous transformation of the ADM mass, 
which is $\delta_\xi Q_\zeta^\mathrm{QL}$
with $\epsilon_\zeta(t) = 1 $. 
For the metric \eqref{GeneralNH} and 
the asymptotic symmetry group \eqref{AsymptKillingR}, 
we obtain 
\begin{align}
 \delta_\xi Q_{\partial_{\hat t}}^\mathrm{QL} 
 &= 
 \frac{1}{8\pi G_N}\int d\phi\,d\psi\,d\theta 
 \frac{k_i k_j \gamma_{ij}(\theta)\sqrt{\gamma(\theta)f_\theta(\theta)}}
 {2 \Lambda f_0(\theta)} \epsilon_\xi'''(\hat t) \label{QL} \ ,
\end{align}
where we have introduced a cut-off by putting 
the boundary at $\hat r = \Lambda$. 
By using the explicit form of the near horizon metric 
\eqref{NHmetric}, it turns out that 
\begin{align}
 \delta_\xi Q_{\partial_{\hat t}}^\mathrm{QL}
 &=
 \frac{\pi r_0 \mu_0}{4 G_N \Lambda}\epsilon_\xi'''(\hat t) \ . 
\end{align}
From the definition of the near horizon coordinate of $\hat r$, 
it must satisfy 
\begin{equation}
 \hat r \ll \frac{2 r_0}{\sqrt{\mu_0}}\epsilon^{-1} \ , 
\end{equation}
in order for the expansion in $\epsilon$ to be valid. 
Therefore, we put the boundary of the near horizon geometry at 
\begin{equation}
 \Lambda = \frac{2 r_0}{\sqrt{\mu_0}}\epsilon^{-1} \ . \label{CutOff}
\end{equation}
The central charge is related to the anomalous transformation 
of the charge as 
\begin{equation}
 \delta_\xi Q = \frac{c^{(QL)}}{12}\epsilon_\xi'''(\hat t) 
  + (\text{non-anomalous terms}) \ . 
\end{equation}
Then, the central charge can be evaluated as 
\begin{equation}
 c^{(QL)} =  \frac{3\pi\mu_0^{3/2}}{2 G_N}\epsilon \ . 
  \label{QLCentral}
\end{equation}
The Frolov-Thorne temperature for 
right mover can be read off 
from the Boltzmann factor with respect to the charge 
$Q_{\partial_{\hat t}}$,  
and given by 
\begin{equation}
 T = \frac{\sqrt{\hat\mu}}{2\pi}\ . 
\end{equation}
Using the Cardy formula, 
we obtain 
\begin{equation}
 S = \frac{\pi^2}{3}c^{(QL)} T 
  = \frac{\pi^2 \mu_0^{3/2}\sqrt{\hat\mu}}{4G_N}\epsilon\ . 
  \label{CardyR}
\end{equation}
Since the Bekenstein-Hawking entropy is 
expanded in the near-extremal case as 
\begin{equation}
 S = \frac{\pi^2}{2G_N}\mu r_+ = 
  \frac{\pi^2}{2G_N}\mu_0 
  \left(r_0 + \frac{1}{2}\epsilon\sqrt{\mu_0\,\hat\mu}
   +\mathcal O(\epsilon^2)\right) \ , 
\end{equation}
the expression \eqref{CardyR} agrees 
with the leading non-extremal correction. 

Even though the Bekenstein-Hawking entropy is 
correctly reproduced, 
the identification of the cut-off \eqref{CutOff} 
is just a rough estimation. 
In order to justify this choice of the cut-off, 
we compare this result with 
the AdS$_3$/CFT$_2$ correspondence in the next section.

\section{Comparison with the AdS$_3$/CFT$_2$ correspondence}\label{sec:AdS}

When one of the angular momenta vanishes, 
the near horizon geometry of 
the five-dimensional Myers-Perry black hole 
has the structure of AdS$_3$. 
If this momentum is not exactly zero 
but infinitesimally small in the near-extremal case, 
this AdS$_3$ part 
becomes the BTZ black hole. 
In this case, we can simply apply 
the ordinary AdS/CFT correspondence 
for the Myers-Perry black hole. 
In this section, we consider 
such a case and compare it with 
the previous two results. 

In the previous sections, we have considered 
the near-extremal case in which the 
non-extremality is infinitesimally small. 
Here, we assume that one of the angular momenta is also infinitesimally small and 
of the same order to the non-extremality. 
We define the parameters $\tilde\mu$ and $\tilde b$ 
by the following relations, 
\begin{align}
 \mu &= a^2+b^2 + 2 a \epsilon^2 \tilde m \ , & 
 b &= \epsilon^2 \tilde b \ , 
\end{align}
and redefine the near horizon coordinates as 
\begin{align}
 t &= \epsilon^{-1} \tilde t \ , &
 r &= \epsilon \tilde r \ , 
 \\
 \phi &= \tilde \phi + \frac{a}{r_0^2+a^2}t \ , & 
 \psi &= \epsilon^{-1} \tilde \psi\  . 
\end{align}
By taking the near horizon limit $\epsilon\to 0$, 
the metric becomes 
\begin{align}
 ds^2 
 &= 
 -\frac{\cos^2\theta}{a^2}\frac{\tilde\Delta}{\tilde r^2} d\tilde t^2 
 + \frac{a^2 \cos^2\theta\,\tilde r^2}{\tilde \Delta} d\tilde r^2 
 +\cos^2\theta\,\tilde r^2
 \left(
 d\tilde \psi -\frac{b}{\tilde r^2}d\tilde t
 \right)^2 
 \notag\\
 &\quad
 +a^2\frac{\sin^2\theta}{\cos^2\theta}d\tilde\phi^2 
 + a^2\cos^2\theta\,d\theta , \label{NHAdS}
\end{align}
where $\tilde\Delta$ is given by 
\begin{align}
 \tilde \Delta &= \tilde r^4 - 2 a \tilde m \tilde r^2 + a^2 \tilde b^2 
 = (\tilde r^2 - \tilde r_+^2)(\tilde r^2 - \tilde r_-^2) \ , 
\end{align}
and $\tilde r_\pm$ is the 
positions of the outer and inner horizons in terms of $\tilde r$, 
which is expressed as 
\begin{equation}
 \tilde r_\pm^2 = \epsilon^{-2} r_\pm^2 
  = a \left(\tilde m \pm \sqrt{\tilde m^2 - \tilde b^2}\right) \ . 
\end{equation}
Then, this geometry has the structure of the BTZ black hole. 
Strictly speaking, $\tilde\psi$ has an infinitesimal period $2\pi\epsilon$. 

The analysis of the asymptotic symmetry can be 
applied to this geometry straightforwardly. 
For simplicity, we introduce the light-corn coordinates 
\begin{equation}
 x^\pm = \tilde\psi \pm \frac{\tilde t}{a} \ . 
\end{equation}
By imposing the boundary condition: 
\begin{equation}
 \mathcal O(\chi_{\mu\nu}) = 
  \bordermatrix{
  & x^+ & \tilde r & x^- & \tilde\phi & \theta \cr
  x^+
  & \mathcal O(r^{0}) 
  & \mathcal O(r^{-2})
  & \mathcal O(r^{0}) 
  & \mathcal O(r^{-2}) 
  & \mathcal O(r^{-3}) 
  \cr 
  \tilde r 
  & 
  & \mathcal O(r^{-4})
  & \mathcal O(r^{-2})
  & \mathcal O(r^{-3})
  & \mathcal O(r^{-4}) 
  \cr 
  x^-
  & 
  & 
  & \mathcal O(r^{0})
  & \mathcal O(r^{-2})
  & \mathcal O(r^{-3}) 
  \cr 
  \tilde\phi 
  & 
  & 
  &
  & \mathcal O(r^{-2})
  & \mathcal O(r^{-3})
  \cr
  \theta
  &
  & 
  & 
  & 
  & \mathcal O(r^{-3})
  }\ , \label{AsymptCondAdS}
\end{equation} 
which is the same as the original work by Brown and Henneaux \cite{bh} 
for AdS$_3$ part, 
we obtain the following asymptotic symmetry groups: 
\begin{subequations}\label{KillingAdS}
 \begin{align}
  \xi^{(+)}
  &= 
  \epsilon_+(x^+) \partial_+ 
  - \frac{1}{2} \tilde r \epsilon_+'(x^+) \partial_{\tilde r} 
  - \frac{a^2}{2 \tilde r^2} \epsilon_+''(x^+) \partial_- \ , 
  \\
  \xi^{(-)} 
  &= 
  \epsilon_-(x^-) \partial_- 
  - \frac{1}{2}\tilde r \epsilon_-'(x^-) \partial_{\tilde r} 
  - \frac{a^2}{2 \tilde r^2} \epsilon_-''(x^-) \partial_+ \ .  
 \end{align}
\end{subequations}
The coordinate $x^+$ and $x^-$ parametrize 
the almost same directions to $\hat t$ and $\hat\phi$, 
and hence, these asymptotic symmetry groups are almost equivalent 
to those studied in the previous sections. 
In order to see this, we take the limit of $b\to 0$ 
of the near horizon geometry \eqref{NHmetric}. 
Comparing the definition of the coordinates 
$(\hat t, \hat\psi)$ with $(x^+, x^-)$
we obtain the following relations: 
\begin{align}
 x^+ &= \hat t + \mathcal O(\epsilon) \ , &
 x^- &= \epsilon\hat\psi \ . 
\end{align}
Then, by taking $\epsilon\to 0$ limit with 
these coordinates and $b=\epsilon^2\tilde b$, 
the near horizon metric \eqref{NHmetric} becomes%
\footnote{
Strictly speaking, here we take the near-extremal limit and 
the small $b$ limit separately, 
and hence, 
the parameter $\epsilon$ here and 
that in \eqref{NEcond}, \eqref{NHCoord} 
should be distinguished. 
The metric \eqref{NEAdS} agrees with \eqref{NHAdS} 
only in the near-extremal case, because we first take 
the near-extremal limit for \eqref{NEAdS}. 
} 
\begin{align}
 ds^2 &= \frac{a^2\cos^2\theta}{4}\hat\mu(dx^+)^2 
 + \sqrt{a^3\tilde b}\,\hat r\,dx^+\,dx^- 
 + a\tilde b\cos^2\theta(dx^-)^2 
 \notag\\
 &\quad
 + \frac{a^2\cos^2\theta}{4\hat\Delta}d\hat r^2 
 + a^2\frac{\sin^2\theta}{\cos^2\theta}d\hat\phi^2 
 + a^2\cos^2\theta\,d\theta^2 \ . \label{NEAdS}
\end{align}
Since the parameters $\hat\mu$ and $\tilde m$ are 
related as $\hat\mu/2 = a(\tilde m - \tilde b)$, 
this expression agrees with \eqref{NHAdS} 
in the extremal limit of $\tilde m \to \tilde b$ 
if we identify%
\footnote{
This relation is consistent with
the definitions of $\hat r$ and $\tilde r$, 
since the definition of $\hat r$ can be rewritten as 
$ r^2 = r_0^2 + \epsilon r_0\sqrt{\mu_0}\,\hat r $ 
in $\epsilon\to 0$ limit. 
} 
\begin{equation}
 \tilde r^2 \sim a\tilde b + \sqrt{a^3\tilde b}\,\hat r . 
\end{equation}
Then, excluding the last terms in \eqref{KillingAdS}, 
which are asymptotically subleading contributions, 
we obtain 
\begin{align}
 \xi^{(+)} 
 &\sim 
 \epsilon_+(\hat t) \partial_{\hat t} 
 - \hat r \epsilon_+'(\hat t)\partial_r + \mathcal O(\epsilon) \ , 
 \\
 \xi^{(-)} 
 &\sim 
 \epsilon^{-1}
 \left(
 \hat\epsilon_-(\hat\psi) \partial_{\hat\psi} 
 - \hat r \hat\epsilon_-'(\hat\psi) \partial_{\hat r}
 \right) + \mathcal O(\epsilon^0) \ , 
\end{align}
where $\hat\epsilon_-(\hat\psi) = \epsilon_-(\epsilon\hat\psi)$. 
Therefore, $\xi^{(+)}$ and $\xi^{(-)}$ correspond to 
the asymptotic symmetry groups for 
the right and left movers, respectively. 

These two sets of vectors form the Virasoro algebras 
as in the ordinary AdS$_3$ case, 
but have slightly different structures. 
Since the coordinate $\tilde\psi$ has 
an infinitesimal period of $2\pi\epsilon$, 
light-cone coordinates $x^+$ and $x^-$ must satisfy 
the following periodicities
\begin{align}
 x^+ &\sim x^+ + 2\pi n \epsilon \ , & 
 x^- &\sim x^- + 2\pi n \epsilon \ . 
\end{align}
Then, the functions $\epsilon_+(x^+)$ and $\epsilon_-(x^-)$ 
can be expanded with the following forms: 
\begin{subequations}\label{FunctExp}
 \begin{align}
  \epsilon_+(x^+) &= e^{inx^+/\epsilon} \ , & 
  \epsilon_-(x^-) &= e^{inx^-/\epsilon} \ , 
 \end{align}
\end{subequations}
with arbitrary integers $n$. 
Now we define $\xi_n^{(+)}$ and $\xi_n^{(-)}$ 
by $\xi^{(+)}$ and $\xi^{(-)}$ with \eqref{FunctExp}. 
Then, these vectors form the following algebras: 
\begin{align}
 [\xi_n^{(+)}, \xi_m^{(+)}] 
 &= -i \frac{n-m}{\epsilon} \xi_{n+m}^{(+)} \ , 
 \\
 [\xi_n^{(-)}, \xi_m^{(-)}] 
 &= -i \frac{n-m}{\epsilon} \xi_{n+m}^{(-)} \ , 
\end{align}
respectively. Here we have an additional factor $\epsilon^{-1}$ 
which comes from the period of $\tilde\psi$. 

Before calculating the central charge, we discuss 
the general property of the Virasoro algebra with an
additional factor. 
In general, the following vector forms the Virasoro algebra: 
\begin{equation}
 \xi = f(x) \partial_x - r f'(x) \partial_r \ . 
\end{equation}
If the coordinate $x$ has the period of $2\pi \beta$, 
the function $f(x)$ must respect this periodicity. 
We define $\xi_n$ by $\xi$ with 
\begin{equation}
 f_n(x) = e^{inx/\beta} \ . 
\end{equation}
Then, this vector forms the following algebra 
\begin{equation}
 [\xi_n,\xi_m] = \frac{n-m}{\beta}\xi_{n+m} \ . 
\end{equation}
This is the Virasoro algebra but has 
an additional factor of $1/\beta$. 
This factor appears because the vector $\xi$ 
is not dimensionless and hence the algebra depends on 
the choice of the coordinate $x$. 
The factor $\beta$ can be absorbed by 
taking the coordinate $x$ to have the period of $2\pi$, 
or equivalently, redefining $\xi\to\xi' = \beta\xi$. 

The Noether charges $\mathcal L_n$ associated with these vector can have the central extension,  
and obeys the following algebra: 
\begin{equation}
 [\mathcal L_n,\mathcal L_m] 
  = -i(n-m)\mathcal L_{n+m} 
  -i \delta_{n+m,0} n^3 \frac{c}{12} \ , 
\end{equation}
where we have rescaled the vector such that 
the algebra takes the standard form. 
In the original definition of the asymptotic symmetry groups, 
we have chosen the normalization such that $\xi_0$ 
gives a conjugate momentum of $x$, namely $\xi_0 = \partial_x$. 
Then, the algebra becomes 
\begin{equation}
 [L_n,L_m] = -i\frac{n-m}{\beta}L_{n+m} 
  -i \delta_{n+m,0} n^3 \frac{c^{(cov)}}{12\beta^3} \ . 
  \label{GeneralAlgebra}
\end{equation}
Here, we have chosen the ``covariant'' definition of 
the central charge such that 
the central extension is related to the anomalous transformation 
of the charge as 
\begin{equation}
 \delta Q_{\xi} \sim \frac{c^{(QL)}}{12}f'''(x) \ , 
\end{equation}
where $Q_\xi\sim \sum L_n f_n(x)$. 
Therefore, the central charge we have derived 
in the previous section is not $c$ but $c^{(cov)}$. 
By using this covariant definition, 
the generators $L_n$, central charge $c^{(cov)}$
and the Frolov-Thorne temperature $T$ 
have the following scaling properties: 
\begin{align}
 x &\to \lambda x \ , & 
 L_n &\to \lambda^{-1} L_n \ , \notag\\
 c^{(cov)} &\to \lambda c^{(cov)} \ , & 
 T &\to \lambda^{-1} T \ , \label{ScaleProp}
\end{align}
where the Frolov-Thorne temperature $T$ is the weight 
for the charge $L_0$. 
These behaviors imply that we can also apply 
the Cardy formula in these definitions. 
These two definitions, 
the standard scale-invariant and covariant one, 
are related with each other by 
\begin{align}
 \mathcal L_n &= \beta L_n \ , & 
 c &= \frac{c^{(cov)}}{\beta} \ , & 
 \mathcal T &= \beta T \ , 
\end{align}
where $\mathcal T$ is the scale-invariant 
Frolov-Thorne temperature 
which is the weight for $\mathcal L_0$. 
Obviously the Cardy formula takes the same form 
for both of these two definitions. 
These two sets of definitions are exactly same when the period of 
the coordinate $x$ is $2\pi$. 
%
%

Now, we evaluate the central charge. 
By using the definition of \cite{bb,bc}, 
the central extensions of the algebras for $\xi^{(+)}$ and $\xi^{(-)}$ are obtained as 
\begin{equation}
 \frac{1}{8\pi G_N}
  \int k_{\xi_n^{(\pm)}}[\pounds_{\xi_m^{(\pm)}} \bar g,\bar g] 
  = -i\frac{\pi}{8} \delta_{n+m,0} a^3 \frac{n^3}{\epsilon^2} \ . 
\end{equation}
Since the period of $x^+$ and $x^-$ are $2\pi\epsilon$, 
the scale-covariant central charges 
for $\xi^{(+)}$ and $\xi^{(-)}$ are 
\begin{equation}
 c_\pm^{(cov)} = \epsilon \frac{3\pi a^3}{2 G_N} \ . \label{CentralChargeAdS}
\end{equation}
The Frolov-Thorne temperatures associated with 
$\partial_+$ and $\partial_-$ are given by 
\begin{equation}
 T_\pm = \frac{\tilde r_+ \pm \tilde r_-}{2\pi a} \ . 
\end{equation}
Then, the Cardy formula reproduces the Bekenstein-Hawking entropy
of \eqref{NHAdS}:
\begin{align}
 S_{CFT} &= \frac{\pi^2}{3} c_+^{(cov)} T_+ + \frac{\pi^2}{3} c_-^{(cov)} T_- \notag\\
 &= \frac{\pi^2}{2G_N}a^2 \tilde r_+ \epsilon = S_{BH}\ . 
\end{align}
The central charges \eqref{CentralChargeAdS} are
actually equivalent to those derived in the previous sections. 
Since the coordinates are rescaled as 
\begin{equation}
 \hat\psi \to x^- = \epsilon\hat\psi \ , 
\end{equation}
the central charges before and after the rescaling are related with
each other as
\begin{equation}
 c_-^{(cov)} = \epsilon\, c_{\hat\psi} \ . 
\end{equation}
From \eqref{CentralChargePsi} and \eqref{CentralChargeAdS}, 
it is clear that $c_{\hat\psi}$ and $c_-^{(cov)}$ satisfy this relation 
in $b\to 0$ limit. 
For the right mover, the coordinates $\hat t$ and $x^+$ 
are equivalent in $\epsilon\to 0$ limit. 
In fact, the central charge $c^{(QL)}$ in \eqref{QLCentral} 
equals to $c_+^{(cov)}$ in $b\to 0$ limit. 
This justifies the identification of the cut-off \eqref{CutOff}. 

Even though the asymptotic symmetry groups 
\eqref{AsymptKillingLBoth} and \eqref{AsymptKillingR}
agree with those in AdS$_3$ at the leading order of $\epsilon$, 
there are higher order corrections. 
The near horizon coordinate $\hat t$ is the time coordinate
in AdS$_3$ and not 
exactly equivalent to 
the light-cone coordinate $x^+$.
We can also introduce an asymptotic symmetry group
of the time direction for AdS$_3$ 
by imposing a suitable boundary condition
. 
However it is rather natural to 
define a new near horizon limit to obtain asymptotic symmetry groups
for the general extremal Myers-Perry black holes. 
We will discuss it in the next section.

\section{New near horizon limit of 5D Myers-Perry black hole}\label{sec:NewNH}

In this section, we consider 
another definition of the near horizon limit. 
We have defined the near horizon coordinates by \eqref{NHCoord}. 
However, the central charge of right mover 
cannot be calculated by the covariant phase space method given by \cite{bb,bc}, 
and hence we used the quasi-local charge \cite{by,bk}. 
In the new coordinates, the asymptotic Virasoro symmetries
are realized along the light-cone coordinates $(x^+,x^-)$ similarly 
to the AdS$_3$ spacetime, while those in the usual coordinates
are associated with $(\hat t,\hat\psi)$ directions. 
We will show that the definition of \cite{bb,bc} 
also gives the central charge and reproduces 
the entropy via the Cardy formula. 

We define the new coordinates $x^+$ and $x^-$, 
and redefine $\hat r$ as follows: 
\begin{align}
 x^+ &= \epsilon
 \left(\psi+\frac{a-b}{\mu}t\right) \ , & 
 x^- &= \psi - \frac{a+b}{\mu}t \ , & 
 \hat r &= r_0 + \epsilon \frac{a}{2}\hat r \ . 
\end{align}
In the near-extremal case of \eqref{NEcond}, 
the near horizon geometry 
takes the same form as \eqref{NHmetric}, 
but the coordinates $\hat t$ and $\hat\psi$ 
are replaced with $x^+$ and $x^-$. 
Namely, taking $\epsilon\to 0$ limit, we obtain 
\begin{align}
 ds^2 
 &= 
 - \frac{\rho_0^2\hat\Delta}{4} (dx^+)^2 
 + \frac{\rho_0^2}{4\hat\Delta}d\hat r^2 
 + \rho_0^2 d\theta^2 
 \notag\\
 &\quad
 +\frac{a^2\mu_0\sin^2\theta}{\rho_0^2}
 \left(
 d\hat\phi + k_{\hat\phi}\hat r dx^+
 \right)^2
 +\frac{b^2\mu_0\cos^2\theta}{\rho_0^2}
 \left(
 dx^- + k_{\hat\psi}\hat r dx^+
 \right)^2
 \notag\\
 &\quad 
 +\frac{\mu_0r_0^2}{\rho_0^2}
 \left[
 \sin^2\theta\left(
 d\hat\phi + k_{\hat\phi}\hat r dx^+
 \right)
 +\cos^2\theta\left(
 dx^- + k_{\hat\psi}\hat r dx^+
 \right)
 \right]^2 \ , \label{NewNHmetric}
\end{align}
where we have also redefined $\hat\Delta$ by 
\begin{equation}
 \hat\Delta = \hat r^2 - \frac{\mu_0}{a^2}\hat\mu . 
\end{equation}
Then, we can obtain the asymptotic symmetry group 
in the same fashion as the previous section 
(but the coordinate $\hat t$ and $\hat\psi$ 
are replaced with $x^+$ and $x^-$). 
By using the same boundary condition as \eqref{AsymptCondR}, 
we obtain the asymptotic symmetry groups, 
\begin{align}
 \xi &= 
  \Big(
   \epsilon_+(x^+) + \frac{\epsilon_\xi''(x^+)}{2\hat r^2}
  \Big)
  \partial_{+} 
  + 
  \Big(
   - \hat r \epsilon_\xi'(x^+) 
 + \frac{\epsilon_\xi'''(x^+)}{2\hat r}
  \Big)
  \partial_{\hat r} 
 \notag\\
 &\qquad
  + 
  \Big(
   C_\phi - \frac{k_{\hat\phi} \epsilon_\xi''(x^+)}{\hat r}
  \Big)
  \partial_{\hat\phi} 
  + 
  \Big(
   C_{\hat\psi} - \frac{k_{\hat\psi} \epsilon_\xi''(x^+)}{\hat r}
  \Big)
  \partial_{-} 
  + 
  \mathcal O(\hat r^{-3}) \ .  \label{NewAsymptKilling}
\end{align}
Since $\psi \sim \psi + 2\pi$, 
the coordinates $x^+$ and $x^-$ 
have the periodicity of 
\begin{align}
 x^+ &\sim x^+ + 2\pi n \epsilon \ , & 
 x^- &\sim x^- + 2\pi n \ . 
\end{align}
Then, the function $\epsilon_\xi(x^+)$ must have 
the following form: 
\begin{equation}
 \epsilon_\xi(x^+) = e^{inx^+/\epsilon} \ . 
  \label{Funct+}
\end{equation}
We define $\xi_n$ by $\xi$ with \eqref{Funct+}. 
Then, $\xi_n$ forms the following algebra: 
\begin{equation}
 [\xi_n,\xi_m] = -i\frac{n-m}{\epsilon}\xi_{n+m} \ . 
\end{equation}

Now let us consider the conserved charge defined by \eqref{conservedcharge}
in our new coordinates.
In the previous definition of the near horizon coordinates, 
$\hat t$ is equivalent to the time of the original coordinates 
up to the scaling factor. 
Hence, the timeslice in the near horizon geometry 
is also defined on $\hat t = $ const.\ plane. 
However, in the new definition, 
the coordinate $x^+$ is not equivalent to 
the original time. 
We should perform the integration 
on the original timeslice, hence 
the timeslice $\Sigma$ (and its boundary $\partial\Sigma$) 
is not $x^+ = $ const.\ plane. 

For the asymptotic symmetry groups \eqref{NewAsymptKilling}, 
we obtain 
\begin{align}
 \tilde k_\zeta^{+r}[\pounds_\xi\bar g,\bar g] &= 0 \ , &
 \tilde k_\zeta^{-r}[\pounds_\xi\bar g,\bar g] 
 &= -\frac{k_{\hat\psi}}{f_0(\theta)}
 \epsilon_\zeta(x^+)\epsilon_\xi'''(x^+) \ ,  
\end{align}
where $f_0(\theta)$, $k_\psi$, etc.\ are the same as
those in the previous section. 
Since the central extension becomes 
\begin{equation}
 \frac{1}{8\pi G_N}\int_{\partial\Sigma} 
  k_{\xi_m}[\pounds_{\xi_n}\bar g, \bar g] = 
  \frac{\pi k_{\hat\psi}}{2G_N \epsilon^2} 
  \int d\theta\sqrt{\gamma(\theta) f_\theta(\theta)} \ ,  
\end{equation}
the scale-covariant central charge is evaluated as 
\begin{equation}
 c_{+}^{(cov)} = \frac{6 \pi k_{\hat\psi}\epsilon}{G_N} 
  \int d\theta\sqrt{\gamma(\theta) f_\theta(\theta)} \ .  
\end{equation}
Using the explicit form of the metric \eqref{NewNHmetric}, 
we obtain 
\begin{equation}
 c_+^{(cov)} = \frac{3\pi a\mu_0}{2 G_N} \epsilon \ . \label{Cplus}
\end{equation}
The central charge for left mover can be calculated straightforwardly. 
Since the additional term gives only 
the $\mathcal O(\epsilon)$ corrections, 
the central charge for left mover 
equals to \eqref{CentralChargePsi}. 
It should be noted that the scale-invariant central charge is 
given by $c_+ =c_+^{(cov)}/\epsilon $, and equals to 
the value for left mover $c_- = c_{\hat\psi}$. 

The Frolov-Thorne temperatures 
associated with $\partial_+$ and $\partial_-$ are 
given by 
\begin{align}
 T_+ &= \epsilon^{-1} 
 \frac{r_+ - r_-}{2\pi a} 
 \to \frac{\sqrt{\mu_0\hat\mu}}{2\pi a} \ , 
 \\
 T_- &= \frac{r_+ + r_-}{2\pi a} 
 \to \frac{b}{\pi} \ . 
\end{align}
Then the Cardy formula reproduces 
the Bekenstein-Hawking entropy up to ${\cal O} (\epsilon^2)$: 
\begin{align}
 S &= \frac{\pi^2}{3} c_+^{(cov)} T_+ + \frac{\pi^2}{3} c_-^{(cov)} T_- \notag\\
  &= \frac{\pi^2}{2G_N}\mu_0 
  \left(r_0 + \frac{1}{2}\epsilon\sqrt{\mu_0\,\hat\mu}
   +\mathcal O(\epsilon^2)\right) \ .  
\end{align}

Before closing this section, 
we would like to comment on the definition of 
the new near horizon coordinates. 
First, the definitions of $x^\pm$
is the same as those of the 
hidden conformal symmetry \cite{Krishnan} 
in the near-extremal limit. 
When one of the angular velocities ($b$) is very small, 
these coordinates, $x^\pm$, are equivalent to 
the light-cone coordinates $x^+$ and $x^-$ 
defined in the previous section
up to the factor of $\epsilon$ for $x^-$. 
In this limit, the central charges $c_\pm^{(cov)}$ 
and the Frolov-Thorne temperatures agree with 
those in Section~\ref{sec:AdS} 
when we take into account the factor of $\epsilon$ 
or use the scale-invariant definitions. 
Since the coordinate $x^+$ and $\hat t$ are related 
as $x^+ = \frac{a}{a+b}\hat t + \mathcal O(\epsilon)$, 
the central charge \eqref{QLCentral} can be 
reproduced from $c_+^{(cov)}$ by using  \eqref{ScaleProp}.

\section{Kerr/CFT revisited}\label{sec:Kerr}

In the previous section, 
we have defined the new near horizon limit 
of the five-dimensional Myers-Perry black hole. 
By using this, 
the central charges for right mover 
can be calculated by using the definition of \cite{bb,bc}. 
In this section, we consider such a new near horizon limit for the four-dimensional Kerr black hole. 

By using the Boyer-Lindquist coordinates, 
the Kerr geometry can be expressed as 
\begin{align}
 ds^2 &= -\frac{\Delta}{\rho^2}(dt-a\sin^2\theta\,d\phi)^2 
 + \frac{\sin^2\theta}{\rho^2}
 \left[ (r^2+a^2)d\phi-a dt \right]^2
 +\frac{\rho^2}{\Delta}dr^2 
 + \rho^2 d\theta^2 \ , 
\end{align}
where $\Delta$ and $\rho^2$ are given by 
\begin{align}
 \Delta &= r^2 -2Mr + a^2 \ , & 
 \rho^2 &= r^2 + a^2 \cos^2\theta \ . 
\end{align}
The Kerr geometry are characterized by two parameters $M$ and $a$ which 
are related to the ADM mass and angular momentum as 
\begin{align}
 M_\mathrm{ADM} &= \frac{M}{G_N} \ , & 
 J &= \frac{a M}{G_N} \ . 
\end{align}
The inner and outer horizons are given by 
\begin{equation}
 r_\pm = M \pm \sqrt{M^2 - a^2} \ ,  
\end{equation}
and the angular velocity at the outer horizon $r_+$ is 
\begin{equation}
 \Omega_H = \frac{a}{r_+^2+a^2} \ . 
\end{equation}
The Hawking temperature and the Bekenstein-Hawking entropy 
are given by 
\begin{align}
 T_H &= \frac{r_+-r_-}{4\pi M r_+} \ , & 
 S &= \frac{2\pi M r_+}{G_N}\ . 
\end{align}

We consider the near-extremal case, 
and define a non-extremality parameter $\hat r_H$ as 
\begin{equation}
 M = a (1+\epsilon^2\frac{\hat r_H^2}{2}) \ .  
\end{equation}
Then, the geometry is parametrized by $a$ and $\hat r_H$.
New near horizon coordinates 
$x^\pm$ and $\hat r$ are defined by the following relations, 
\begin{align}
 x^+ &= \epsilon\phi \ , &
 x^- &= \phi - \frac{a\,t}{2M^2} \ , &
 r &= a (1+\epsilon \hat r) \ . \label{NHKerr}
\end{align}
Here, the combination of $t$ and $\phi$ in 
the definition of $x^+$ and $x^-$ are 
the same as those appeared in analysis of the 
hidden conformal symmetry \cite{cms} (in the definition of $w^\pm$). 
 The geometry is expressed in the near horizon limit of $\epsilon\to 0$ as 
\begin{align}
 ds^2 &= -(\hat r^2 - \hat r_H^2) f_0(\theta) (dx^+)^2 
 + f_\phi(\theta) (dx^- + \hat r dx^+)^2 
 \notag\\
 &\quad 
 + f_0(\theta) \frac{d\hat r^2}{\hat r^2 -\hat r_H^2} 
 + f_0(\theta) d\theta^2 \ , \label{NHmetricKerr}
\end{align}
where $f_0(\theta)$ and $f_\phi(\theta)$ are given by 
\begin{align}
 f_0(\theta) &= a^2 (1+\cos^2\theta) \ , & 
 f_\phi(\theta) &= \frac{4a^2\sin^2\theta}{1+\cos^2\theta} \ . 
\end{align}
This geometry has the same form to the ordinary 
so-called NHEK geometry, 
but $t$ and $\phi$ of the near horizon coordinates 
are replaced with $x^+$ and $x^-$, respectively. 

The boundary condition for right mover 
is obtained in \cite{mty}. 
We use this boundary condition by replacing 
$\hat t$ and $\hat\phi$ with $x^+$ and $x^-$:  
\begin{equation}
 \mathcal O(\chi_{\mu\nu}) = 
  \bordermatrix{
  & x^+ & \hat r & x^- & \theta \cr
  x^+ 
  & \mathcal O(r^{0}) 
  & \mathcal O(r^{-3})
  & \mathcal O(r^{-2}) 
  & \mathcal O(r^{-3}) 
  \cr 
  \hat r 
  & 
  & \mathcal O(r^{-4})
  & \mathcal O(r^{-3})
  & \mathcal O(r^{-4})
  \cr 
  x^- 
  & 
  & 
  & \mathcal O(r^{-2})
  & \mathcal O(r^{-3})
  \cr 
  \theta
  & 
  & 
  & 
  & \mathcal O(r^{-3})
  }\ . 
\end{equation}
Then, the following asymptotic symmetry group 
satisfies this boundary condition: 
\begin{align}
 \xi &= 
  \Big(
   \epsilon_\xi(x^+) + \frac{\epsilon_\xi''(x^+)}{2\hat r^2}
  \Big)
  \partial_+
  + 
  \Big(
   - \hat r \epsilon_\xi'(x^+) + \frac{\epsilon_\xi'''(x^+)}{2\hat r}
  \Big)
  \partial_{\hat r} 
 \notag\\
 &\quad
  + 
  \Big(
   C - \frac{\epsilon_\xi''(x^+)}{\hat r}
  \Big)
  \partial_- 
  + 
  \mathcal O(\hat r^{-3})\ .  
\end{align}
Since the period of $\phi$ is $2\pi$, 
the coordinates $x^+$ and $x^-$ have 
the following periodicity: 
\begin{align}
 x^+ &\sim x^+ + 2\pi n \epsilon \ , & 
 x^- &\sim x^- + 2\pi n \ . 
\end{align}
%
The central extension is given by an integration 
of two-from $k_\xi[\pounds_\xi\bar g,\bar g]$ on a timeslice. 
As we have discussed in the previous section, 
the charge should be defined by 
the integration on the timeslice of the 
original coordinates. 
As in the case of the five-dimensional Myers-Perry black hole, 
the near horizon geometry is usually taken 
such that the time direction of the near horizon coordinates 
is equivalent to that of the original one, up to a constant factor. 
In this case, only $\tilde k_\xi^{tr}$ component contributes to the charge 
and the central charge becomes zero \cite{mty}. 
However, in the new near horizon coordinate \eqref{NHKerr}, 
the coordinate $x^+$ is not the time direction of original coordinates. 
Then, $\tilde k_\xi^{-r}$ also contributes to the central extension 
and we obtain 
\begin{equation}
 \frac{1}{8\pi G_N} \int_{\partial\Sigma} 
  k_{\xi_m}^{-r}[\pounds_{\xi_n}\bar g, \bar g] 
  = \delta_{n+m,0}\,n^3 \frac{a^2}{\epsilon^2} \ .
\end{equation}
Then, the scale-covariant central charge is 
\begin{equation}
 c_+^{(cov)} = \frac{12 a^2}{G_N} \epsilon \ . 
\end{equation}
This result agrees with that obtained in \cite{mty}. 
It should be noted that this is the expected value 
of $c_R \equiv c_+ = 12J$ if we use the scale-invariant definition. 

The Frolov-Thorne temperatures associated to 
$\partial_+$ and $\partial_-$ are given by 
\begin{align}
 T_+ &= \epsilon^{-1}\frac{r_+-r_-}{4\pi a} 
 \to \frac{\hat r_H}{2\pi} \ ,
 \\
 T_- &= \frac{r_++r_-}{4\pi a} \to \frac{1}{2\pi}\ . 
\end{align}
Then the Cardy formula gives the entropy 
\begin{equation}
 S = \frac{2\pi a^2}{G_N} \left(a+\epsilon\hat r_H\right) \ , 
\end{equation}
which agrees with the Bekenstein-Hawking entropy up to ${\cal O}(\epsilon^2)$,  
\begin{equation}
 S = \frac{2\pi M r_+}{G_N} 
  = \frac{2\pi a^2}{G_N} 
  \left(a+\epsilon\hat r_H + \mathcal O(\epsilon^2)\right) \ . 
\end{equation}

\section{Conclusion and discussions}\label{sec:Concl}

In this paper, we introduced a new near horizon limit. 
In this limit, structure of the near horizon geometry is the same as
that introduced in \cite{baho}, while 
the original time direction is embedded in a different way. 
Then, conserved charges are slightly modified 
since the timeslice is different in our limit and that in \cite{baho}. 
This limit is useful to describe 
the right mover in the Kerr/CFT correspondence, 
and the central charge can be calculated explicitly. 

By using our new definition of the near horizon limit, 
the charge density depends on the angular coordinates 
and hence we can define generators 
simply integrating on the timeslice. 
The central charge can be calculated by using the 
definition of \cite{bb,bc}, and does not have 
ambiguities of the cut-off. 
It turns out that the Virasoro algebra does not have 
the standard form and depends on the definition of the coordinate. 
By redefining the generators to have the standard algebraic relation, 
the central charge becomes finite and 
satisfies the expected relation of $c_L = c_R = 12J$. 

In our new near horizon coordinates, 
the combinations of coordinates are equivalent to those 
in analyses of the hidden conformal symmetry. 
The presence of the hidden conformal symmetry 
implies the decoupling of the right and left movers. 
In order to apply the Cardy formula separately, 
the right and left movers should be decoupled. 
Hence it is natural that 
the appropriate choice of the coordinates 
is equivalent to those for the hidden conformal symmetry. 

Even though the scale-invariant central charge is finite, 
it is natural to use the covariant definition for the temperature. 
Then, the (covariant) central charge for right mover takes 
an infinitesimally small value, 
and hence, the right mover gives 
subleading contributions in the near-extremal limit.%
\footnote{
If we use the scale-invariant definition, 
the Frolov-Thorne temperature becomes infinitesimally small. 
Therefore, the right mover contributes to the subleading corrections, 
independent to the definition of the central charge.  
} 
We took the near-extremal limit and 
considered its leading corrections. 
However, we did not include all the next-to-leading 
contributions in this limit. 
Hence, we cannot exclude the possibility 
that these contributions affect 
the near-extremal corrections. 
In order to see this, 
we have to study subleading corrections, or 
consider more general non-extremal cases. 

We computed the central charges for each sector separately 
by giving each boundary condition. 
Although we reproduce the Bekenstein-Hawking entropy 
by summing up the entropy of each sector, it is desirable 
to find a boundary condition that admit two Virasoro algebras 
as asymptotic symmetry groups. 
One example is given by \cite{mty2}, but it cannot fix 
higher order corrections of 
the asymptotic symmetries for right mover 
which contribute to the central charges. 
Investigation in this direction will give further evidence for 
the Kerr/CFT correspondence. 

\vspace{1cm}
 \centerline{\bf Acknowledgements}
 We are grateful to T. Hartman and N. Matsumiya for valuable discussions,
 and N. Matsumiya for collaboration on a new near horizon limit at an earlier stage.
 TN would like to thank all members of the High Energy Physics Theory Group of 
 the University of Tokyo for hospitality during his stay.
 The work of TN was supported in part by the US NSF under Grants No.\,PHY-0844827 and
PHY-0756966.

\appendix 

\section{Counter terms for the quasi-local charge}\label{sec:CT}

In this appendix, we consider counter terms 
for the quasi-local charges. 
The quasi-local charge is defined by \eqref{QLCharge}, and 
we have defined the regularized 
surface energy-momentum tensor $\tau_{\mu\nu}$ 
by \eqref{SurfaceEMTensor}. 
Instead of taking difference from 
the background in \eqref{SurfaceEMTensor}, 
a counter term can be introduced to regularize 
the surface energy-momentum tensor. 
In this case, any covariant counter terms 
cannot terminate all of divergent terms in 
the surface energy-momentum tensor. 
Here, we take the counter term 
such that $Q_\xi^\mathrm{QL}$ becomes finite. 
Then, the surface energy-momentum tensor is given by 
\begin{equation}
 \tau^{\mu\nu} = T^{\mu\nu} 
  + \lambda g^{\mu\nu} \ , \label{EMwithCT}
\end{equation}
where $g_{\mu\nu}$ is the induced metric on the boundary and 
$\lambda$ is a constant. 
Using this definition of $\tau_{\mu\nu}$, 
the quasi-local charge for \eqref{GeneralNH} 
has the following divergent terms: 
\begin{align}
 Q_{\xi}^\mathrm{QL} 
 &= 
 -\frac{1}{8\pi G_N}\int d\phi\,d\psi\,d\theta 
 \frac{\Lambda k_i k_j \gamma_{ij}(\theta)\sqrt{\gamma(\theta)f_\theta(\theta)}}
 {2 f_0(\theta)}
 \notag\\
 &\quad 
 + \frac{\lambda}{8\pi G_N}\int d\phi\,d\psi\,d\theta\, 
 \Lambda \sqrt{f_0(\theta)\gamma(\theta)f_\theta(\theta)} 
 + \mathcal O(\Lambda^0) \ . 
\end{align}
The constant $\lambda$ is chosen such that 
these two terms cancel each other. 
For the near horizon geometry of Myers-Perry black hole 
\eqref{NHmetric}, it turns out that 
\begin{equation}
 \lambda = \frac{(a^{3/2}-b^{3/2})\sqrt{a+b}}{3(a-b)} \ . 
\end{equation}
By using this condition, the central extension becomes 
\begin{align}
 \delta_\xi Q_{\partial_{\hat t}}^\mathrm{QL} 
 &= 
 \frac{\lambda}{8\pi G_N}\int d\phi\,d\psi\,d\theta 
 \frac{\sqrt{f_0(\theta)\gamma(\theta)f_\theta(\theta)}}
 {\Lambda} \epsilon_\xi'''(\hat t) 
 \notag\\
 &= \frac{1}{8\pi G_N}\int d\phi\,d\psi\,d\theta 
 \frac{k_i k_j \gamma_{ij}(\theta)\sqrt{\gamma(\theta)f_\theta(\theta)}}
 {2 \Lambda f_0(\theta)} \epsilon_\xi'''(\hat t) \ , 
\end{align}
and then, we obtain the same result to \eqref{QL}. 
If we allow the coefficient $\lambda$ to 
have $\theta$-dependence, 
we can make the charge density to be finite. 
In this case, the coefficient of the counter term becomes 
\begin{equation}
 \lambda(\theta) = \frac{k_i k_j \gamma_{ij}(\theta)}{2} 
  \left(\frac{1}{f_0(\theta)}\right)^{3/2} \ , \label{CTtheta}
\end{equation}
where $f_0(\theta)$, $\gamma_{ij}(\theta)$ and $k_i$ 
just specify the $\theta$-dependence but 
do not respond to the variation with respect to the metric. 
In the limit of $b\to 0$, this coefficient 
can be expanded as 
\begin{equation}
 \lambda(\theta) \to \frac{1}{a\cos\theta} \ . \label{CTlimit}
\end{equation}

This agrees with the counter term introduced in \cite{bk}. 
For small $b$, the near horizon geometry 
has structure of AdS$_3$, whose 
effective radius at each point of $\theta$ is given by 
\begin{equation}
 R = a\cos\theta \ . 
\end{equation}
then, the coefficient of the counter term for this AdS$_3$ is 
\begin{equation}
 \lambda = \frac{1}{R} = \frac{1}{a\cos\theta} \ ,  
\end{equation}
which is equals to \eqref{CTlimit}.

\end{document}